# Artificial Greenhouse Gases as Exoplanet Technosignatures

Edward W. Schwieterman[1,2], Thomas J. Fauchez[3,4,5], Jacob Haqq-Misra[2], Ravi K. Kopparapu[3,4], Daniel Angerhausen[2,6,7], Daria Pidhorodetska[1], Michaela Leung[1], Evan L. Sneed[1], and Elsa Ducrot[8,9]
[1] Department of Earth and Planetary Sciences, University of California, Riverside, CA, USA; eschwiet@ucr.edu
[2] Blue Marble Space Institute of Science, Seattle, WA, USA
[3] NASA Goddard Space Flight Center, 8800 Greenbelt Road, Greenbelt, MD 20771, USA
[4] Sellers Exoplanet Environment Collaboration (SEEC), NASA Goddard Space Flight Center, USA
[5] Integrated Space Science and Technology Institute, Department of Physics, American University, Washington DC, USA
[6] ETH Zurich, Institute for Particle Physics & Astrophysics, Wolfgang-Pauli-Str. 27, 8093 Zurich, Switzerland
[7] National Center of Competence in Research PlanetS, Gesellschaftsstrasse 6, 3012 Bern, Switzerland
[8] LESIA, Observatoire de Paris, CNRS, Université Paris Diderot, Université Pierre et Marie Curie, 5 place Jules Janssen, 92190 Meudon, France
[9] AIM, CEA, CNRS, Université Paris-Saclay, Université de Paris, F-91191 Gif-sur-Yvette, France
Received 2024 January 4; revised 2024 April 24; accepted 2024 May 15; published 2024 June 25

## Abstract

Atmospheric pollutants such as chlorofluorocarbons and $NO_2$ have been proposed as potential remotely detectable atmospheric technosignature gases. Here we investigate the potential for artificial greenhouse gases including $CF_4$, $C_2F_6$, $C_3F_8$, $SF_6$, and $NF_3$ to generate detectable atmospheric signatures. In contrast to passive incidental by-products of industrial processes, artificial greenhouse gases would represent an intentional effort to change the climate of a planet with long-lived, low-toxicity gases and would possess low false positive potential. An extraterrestrial civilization may be motivated to undertake such an effort to arrest a predicted snowball state on their home world or to terraform an otherwise uninhabitable terrestrial planet within their system. Because artificial greenhouse gases strongly absorb in the thermal mid-infrared window of temperate atmospheres, a terraformed planet will logically possess strong absorption features from these gases at mid-infrared wavelengths (~8–12 $\mu$m), possibly accompanied by diagnostic features in the near-infrared. As a proof of concept, we calculate the needed observation time to detect 1 [10](100) ppm of $C_2F_6/C_3F_8/SF_6$ on TRAPPIST-1 f with JWST MIRI's Low Resolution Spectrometer (LRS) and NIRSpec. We find that a combination of 1[10](100) ppm each of $C_2F_6$, $C_3F_8$, and $SF_6$ can be detected with a signal-to-noise ratio $\geq 5$ in as few as 25[10](5) transits with MIRI/LRS. We further explore mid-infrared direct-imaging scenarios with the Large Interferometer for Exoplanets mission concept and find these gases are more detectable than standard biosignatures at these concentrations. Consequently, artificial greenhouse gases can be readily detected (or excluded) during normal planetary characterization observations with no additional overhead.

*Unified Astronomy Thesaurus concepts:* Technosignatures (2128); Exoplanet atmospheres (487); Exoplanets (498); Astrobiology (74); Habitable planets (695); Biosignatures (2018); Greenhouse gases (684); Search for extraterrestrial intelligence (2127)

## 1. Introduction

Detecting life beyond Earth is one of the foundational goals of astrobiology. The specific search for remote indicators of life on exoplanets is a major driver for the design of near to intermediate future space-based mission concepts (National Academies of Sciences, Engineering, & Medicine & others 2019, 2018, 2023; LUVOIR-Team 2019; Gaudi et al. 2020). Biosignatures are spectroscopic or temporal indications that a planet is inhabited with simple (nontechnological) life. Common examples of biosignatures include $O_2$, $O_3$, $CH_4$, $N_2O$, and the vegetation red-edge (Seager et al. 2012; Kaltenegger 2017; Schwieterman et al. 2018), although biosignature interpretations are strongly dependent on planetary context in order to avoid the potential for false positives (e.g., Meadows et al. 2018). In contrast, technosignatures are observational indications of technology that can be detected via astronomical means, which may encompass technological civilizations or the technology left behind by biological or artificial intelligence (Tarter 2007; Haqq-Misra et al. 2022b). Technosignatures may include radio transmissions (Cocconi & Morrison 1959; Tarter 2001; Worden et al. 2017), laser pulses (Stone et al. 2005; Howard et al. 2007; Vides et al. 2019), megastructures (Dyson 1960; Shkadov 1988; Wright 2020), or modifications to the surface or atmosphere of an inhabited (or noninhabited) planet such as solar panels (Lingam & Loeb 2017), city lights (Beatty 2022), or artificial atmospheric gases (Lin et al. 2014; Haqq-Misra et al. 2022a; Seager et al. 2023). See Haqq-Misra et al. (2022b) for a recent review of the search for technosignatures with current and future missions.

Because of the immense distances involved, the search for exoplanetary life is necessarily a search for global biospheres where native life has left a substantial impact on the atmosphere or surface of the planet. The science of exoplanet biosignatures has thus developed to assess candidate planetary biosignatures based on their intrinsic detectability (i.e., absorption properties), ability to accumulate to high concentrations in the atmosphere (i.e., robustness to photochemical reactions that destroy them), and their separability from abiotic planetary processes that would confound interpretations of biogenicity (e.g., Meadows et al. 2018; Schwieterman et al. 2018). The near future potential for characterizing terrestrial exoplanetary atmospheres and surfaces in search of remote biosignatures has opened up the possibility for commensal searches for planetary technosignatures with no added cost (Lingam & Loeb 2019). Commensal search programs are already common within SETI, especially when conducting radio-based observations, e.g., SERENDIP (Bowyer et al. 1988; Sullivan et al. 1997), Allen Telescope Array (DeBoer 2006), MeerKAT (Czech et al. 2021), and COSMIC (Tremblay et al. 2024). In a similar vein to the search for biosignatures, planetary technosignatures (technosignatures limited to the planetary scale, excluding, e.g., stellar megastructures) are functionally a search for "technospheres," where the impact of technology has detectably risen above the nontechnological (abiotic and biotic) background level of the planetary environment (Frank et al. 2017; Haqq-Misra et al. 2020).

In contrast to biosignatures, many technosignatures may provide greater specificity (less "false positive" potential), as many putative technosignatures have more limited abiotic formation channels when compared to biosignatures. On the other hand, technosignatures may be short-lived when compared to biosignatures (Sheikh 2020). If we only take our planet as an example, atmospheric biosignatures may have persisted for up to 4 Gyr (Krissansen-Totton et al. 2018b; Olson et al. 2018), while our civilizational technosignatures are, at most, centuries old. Potential atmospheric technosignatures are further limited by practical observational considerations if such targets are restricted to the same systems where biosignature searches are to be conducted. In a volume-limited survey that includes only the nearest systems where angular resolution constraints allow habitable zones to be directly imaged—for example the ∼25 systems recommended by the National Academy of Sciences 2020 Astronomy and Astrophysics Decadal Survey for the next-generation IR/optical/UV surveyor telescope (National Academies of Sciences, Engineering, & Medicine & others 2023)—it would seem vanishingly unlikely to detect atmospheric technosignatures if we take at face value the ratio of the longevity of technosignatures to biosignatures on Earth. However, Wright et al. (2022) argue that this view is based on the untested premise that nontechnological biospheres are more common than technospheres. In contrast to the constraints of simple life, technological life is not necessarily limited to one planetary or stellar system, and moreover, certain technologies could persist over astronomically significant periods of time. We know neither the upper limit nor the average timescale for the longevity of technological societies (not to mention abandoned or automated technology), given our limited perspective of human history. An observational test is therefore necessary before we outright dismiss the possibility that technospheres are sufficiently common to be detectable in the nearby Universe (Haqq-Misra et al. 2020).

Atmospheric pollutants such as chlorofluorocarbons (CFCs) and $NO_2$ have been proposed as potential remotely detectable atmospheric technosignature gases (Lin et al. 2014; Kopparapu et al. 2021; Haqq-Misra et al. 2022a). However, these compounds are incidental products of industrial civilization and have several deleterious impacts on the planetary environment and on the civilization producing them. For example, CFCs promote the destruction of the ozone layer (Solomon 1999) and $NO_2$ causes respiratory toxicity (Elsayed 1994). Recent years have seen the dramatic reduction of CFC (and related hydrofluorocarbons) use on Earth due to the Montreal Protocol and subsequent amendments (Hu et al. 2017; Fahey et al. 2018). Because industrial $NO_2$ is most often produced as a result of combustion, its production would fall precipitously if fossil fuels were to be phased out. It is therefore possible, even likely, that the window to observe these pollutant gases on a planet inhabited by an industrial civilization would be short in geologic terms.

We propose that certain atmospheric technosignatures do not suffer from this longevity problem. In contrast to industrial pollutants, artificial greenhouse ("terraforming") gases would represent an intentional effort to modify a planet's climate, and could persist for the entire remaining history of a civilization or beyond. Terraforming, by definition, requires sufficient modification of the atmosphere to adjust a planet's global energy balance, which correspondingly implies a large spectral signature in the thermal infrared portion of the planet's spectrum, incidentally supporting remote detectability. Maintaining a terraformed atmosphere may also require the consistent and intentional replenishment of the contributing gases. Fortuitously, many such gases tend to have long atmospheric residence times of thousands to tens of thousands of years (e.g., Mühle et al. 2010), which would help make the cost of doing so nonprohibitive. Civilizations may be motivated to introduce highly efficient greenhouse gases to forestall a global ice age on their own home world caused by analogs to Earth's Milankovich cycles (Berger 1988; Haqq-Misra 2014). Alternatively, they may use terraforming gases to make another planet in their home system (or beyond) more suitable for life, as humans have proposed for Mars (e.g., Graham 2004; Marinova et al. 2005; Dicaire et al. 2013; Pałka et al. 2022). For the case of Mars, the idea of mobilizing available $CO_2$ and other volatile inventories as a terraforming strategy appears to be largely infeasible (Jakosky & Edwards 2018), so the use of additional artificial gases would be needed for an effective terraforming strategy.

A handful of previous studies have studied the use of artificial greenhouse gases for terraforming. Marinova et al. (2005) examine the potential efficacy of several artificial greenhouse gases including carbon tetrafluoride ($CF_4$), hexafluoroethane ($C_2F_6$), perfluoropropane ($C_3F_8$), and sulfur hexafluoride ($SF_6$) to warm modern Mars. These fluorine-bearing gases were chosen in part because of their nontoxic nature and their relative inertness compared to chlorine- or bromine-containing greenhouse gases that catalytically destroy ozone. On a per-molecule basis, each of these species is a far more effective greenhouse gas than $CO_2$ or $H_2O$ due to strong and broad absorption features that overlap with the mid-infrared (MIR) window of a habitable exoplanet (∼8–12 $\mu$m; Mühle et al. 2010; Totterdill et al. 2016; Kovács et al. 2017). Marinova et al. (2005) found that $C_3F_8$ has the largest warming potential of any single gas, but that a mixture of these four gases would be more effective than any one gas alone due to their nonoverlapping absorption features. Dicaire et al. (2013) similarly examined the warming potential of $CF_4$, $C_2F_6$, $C_3F_8$, and $SF_6$ using a Martian global climate model.

Haqq-Misra et al. (2022b) suggested that future exoplanet characterization missions should search for these artificial greenhouse gases including perfluorocarbons (PFCs; $C_xF_y$), $SF_6$, and nitrogen trifluoride ($NF_3$; see their Figures 1 and 3). Elowitz (2022) noted that artificial greenhouse gases such as CFCs, PFCs, and other fluorinated gases including $NF_3$ and $SF_6$, would be ideal technosignature candidates. Seager et al. (2023)

argue that the fully fluorinated gases $SF_6$ and $NF_3$ are particularly compelling technosignatures because they are unlikely to be produced biologically, are produced at low-abiotic rates, and possess spectral features that do not overlap significantly with common gases. While the detectability of CFCs with JWST has been studied (Lin et al. [2014](); Haqq-Misra et al. [2022a]()), to our knowledge, no existing study has quantitatively examined the detectability of PFCs or other fully fluorinated gases with JWST or future observatories, particularly in the context of terraformed planets.

Here we examine the potential detectability of $C_2F_6$, $C_3F_8$, $SF_6$, $NF_3$, and $CF_4$ with current and future observatories using a combination of radiative transfer and instrument simulation models as a first-pass "proof of concept" for detecting artificial greenhouse gases and terraformed planets. We produce both synthetic transmission spectra applicable to observations and emission (thermal infrared) spectra. We correspondingly conduct detectability analyses for a TRAPPIST-1 f-like exoplanet (chosen because it is in the outer habitable zone) with JWST and directly imaged terraformed terrestrial outer habitable zone planets with the Large Interferometer for Exoplanets (LIFE) mission concept (Quanz et al. [2022]()). While we do not conduct self-consistent climate calculations, we do examine the detectability of these gases individually and in combination at levels of 1, 10, and 100 ppm in a 1 bar Earth-like atmosphere, consistent with the range explored by Marinova et al. ([2005]()) that would result in appreciable greenhouse warming on a Mars-like planet (∼1–40 K). We apply our results to inform a generalized approach for fingerprinting exoplanetary atmospheres with artificial greenhouse warming, which should produce an anomalous MIR signature possibly accompanied by incidental but diagnostic near-infrared (NIR) signatures, and discuss the implication of these findings for ongoing and future searches for planetary biosignatures and technosignatures.

## 2. Spectral Inputs and Models

Our goal is to demonstrate that at abundances consistent with climate modification, artificial greenhouse gases may be detectable in exoplanetary spectra. As a proof of concept, we examine five technosignature gases with substantial thermal infrared opacity: $CF_4$, $C_2F_6$, $C_3F_8$, $SF_6$, and $NF_3$ at concentrations of 1, 10, and 100 ppm in otherwise Earth-like atmospheres. This range of concentrations has been previously examined in the context of terraforming Mars (Marinova et al. [2005]()). We additionally examine a combination of $C_3F_8$, $C_2F_6$, and $SF_6$ in a range of 1, 10, and 100 ppm because this combination of gases more thoroughly covers the MIR window for habitable planets (∼8–12 $\mu$m; $T_{surf} \sim 290$ K). Our base atmosphere was sourced from the Intercomparison of Radiation Codes in Climate Models (Luther et al. [1988]()) and represents a midlatitude summer Earth profile. The opacities for each gas were sourced from Sharpe et al. ([2004]()) via Kochanov et al. ([2019]()) and are shown in Figure [1](). We note that only cross sections, rather than line-by-line data, are available for these gaseous species and that these data are additionally limited in wavelength coverage and precision. We examine the impact of these limitations and the need for refined opacity data for potential technosignature molecules in the Discussion section.

### 2.1. Planetary Spectrum Generator

To generate synthetic transmission spectra and quantify the detectability of artificial greenhouse gases in our simulated transmission spectra, we use the Planetary Spectrum Generator (PSG; Villanueva et al. [2018](), [2022]()). PSG is a public and versatile radiative transfer tool that can be employed to simulate a wide variety of planetary environments and has been used extensively for simulating terrestrial planets, particularly the TRAPPIST-1 planetary system (e.g., Fauchez et al. [2019](), [2020](); Pidhorodetska et al. [2020](); Suissa et al. [2020](); Cooke et al. [2023](); Ostberg et al. [2023]()). In our proof of concept transit cases, we use PSG to calculate the number of transits necessary to identify each molecule or combination of molecules investigated for a TRAPPIST-1 f test case, assuming stellar and planetary parameters sourced from NASA's Exoplanet Archive and shown in Table [1](). We simulate synthetic transmission spectra observable by JWST's Mid-Infrared (5–12 $\mu$m) Instrument (MIRI; see Wright et al. [2004]()) at a constant resolving power of $R = 100$ and synthetic spectra observable by JWST's Near-Infrared (1.5–5 $\mu$m) Spectrograph (NIRSpec; see Bagnasco et al. [2007]()) at a constant spectral resolution of 0.022 $\mu$m. For the NIR spectra, we note that opacity data are unavailable for $CF_4$, $C_2F_6$, $C_3F_8$, $SF_6$, and $NF_3$ for wavelengths shorter than 1.5 $\mu$m. Our transit calculations assume a cloud layer from 1 to 0.1 bar (i.e., an ∼15 km tropopause) with a mass mixing ratio of $10^{-9}$ kg kg$^{-1}$ of ice, with an arbitrary fixed 1 $\mu$m effective radius. This cloud layer is not fully opaque and leads to a continuum level at 5.5 km.

To estimate the signal-to-noise ratio (S/N) across the NIRSpec prism and MIRI wavelength ranges and the number of transits required to achieve a 5$\sigma$ detection, we proceed with the following, as in Fauchez et al. ([2022]()): (1) we compute the spectrum without the molecule of interest; (2) we compute the spectrum with the molecule of interest; (3) we compute the difference between steps 1 and 2 across the whole instrument range; (4) we compute the S/N by dividing step 3 by the noise for one transit in each spectral interval; (5) we apply an out-of-transit factor of 1.17 to the noise assuming an out-of-transit time equal to 3 times the in-transit time; (6) the S/N of the molecule across the whole instrument range is then computed following Lustig-Yaeger et al. ([2019]()); and (7) from the S/N, the number of transits required to achieve a 5$\sigma$ detection is given as in Fauchez et al. ([2019]()) and Fauchez et al. ([2020]()). The noise model employed within PSG is identical to the one used in Fauchez et al. ([2022]()) where calculations have been benchmarked with the JWST Exposure Time Calculator for both the NIRSpec prism and MIRI's Low Resolution Spectrometer (LRS). For the NIRSpec prism, an $R = 100$ has been assumed with the SUB512S subarray with rapid readout pattern and two groups per integration (0.225 s frame$^{-1}$). For MIRI LRS, $R = 100$ has also been assumed, with the P750L disperser using the SLITLESSPRISM subarray and the FASTR1 readout pattern with 20 groups per integration (0.15 s frame$^{-1}$). The out-of-transit time has been assumed to be 1.5× the in-transit time before ingress, and 1.5× after egress, therefore leading to an out-of-transit baseline 3 times longer than the in-transit time. We uploaded our planetary spectra within PandExo[10] to compute the transit depth uncertainty per spectral interval for one transit as shown in Appendix [A](). The transit depth uncertainty increases spectra of exoplanet atmospheres (e.g., Charnay et al. 2015; Lustig-Yaeger et al. 2023a). SMART uses the HITRAN line lists (Gordon et al. 2022) to calculate the opacities of major molecules using its companion program LBLABC, with the exception of the artificial gas cross-section data as described above. Our emission spectra assume a planet that has an identical surface gravity and radius as the Earth, in contrast to the TRAPPIST-1 f transit scenario described above. We simulate planetary emission from 5–21 μm at 1 cm$^{-1}$ resolution with SMART, but these spectra are down-binned for S/N calculations with LIFESIM as described below. As this is a first-pass look at the potential detectability of artificial greenhouse gases, we neglect cloud cover in our emitted light spectral simulations and assume a surface emissivity of 1.

---
[10] http://exoctk.stsci.edu/pandexo

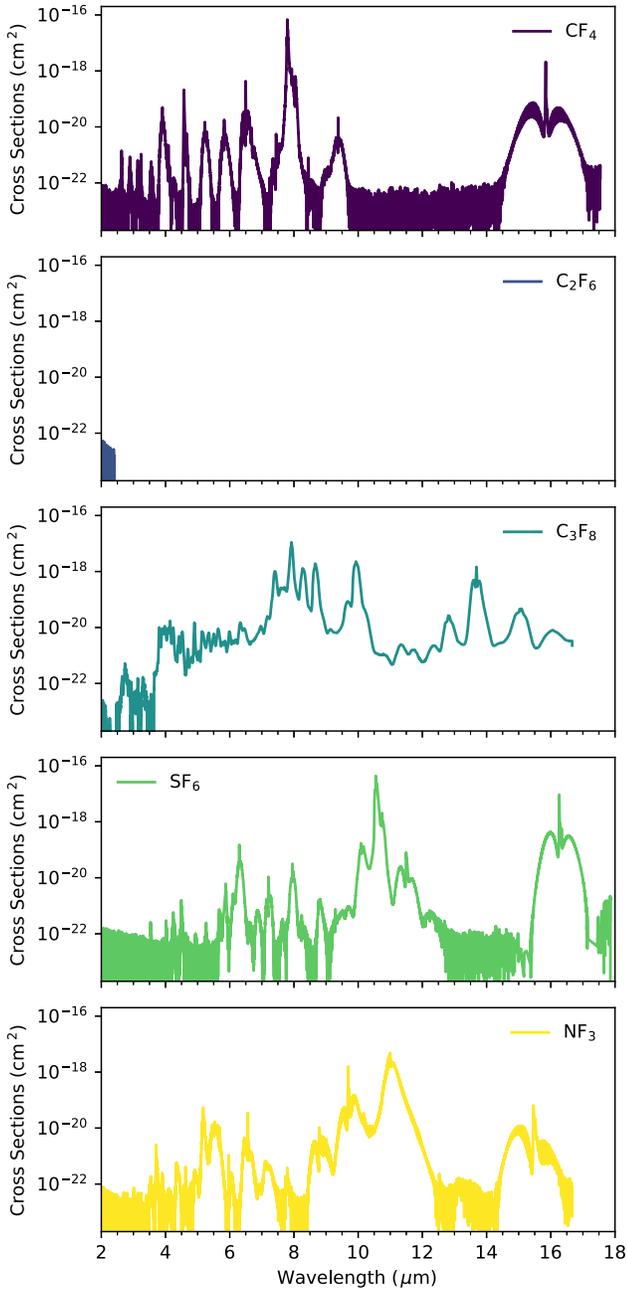

**Figure 1.** Infrared (2–18 μm) cross sections for the atmospheric technosignature molecules examined in this study, which include $CF_4$, $C_2F_6$, $C_3F_8$, $SF_6$, and $NF_3$. Data are sourced from Sharpe et al. (2004) and Kochanov et al. (2019).

**Table 1**
TRAPPIST-1 System Properties Modeled

| Parameter | Value |
| --- | --- |
| TRAPPIST-1 | |
| Distance (pc) | 12.4 |
| Spectral type | M8V |
| $T_{\rm eff}$ (K) | 2566 |
| Mass ($M_\odot$) | 0.0898 |
| Radius ($R_\odot$) | 0.12 |
| TRAPPIST-1 f | |
| Semimajor axis (au) | 0.03849 |
| Mass ($M_\oplus$) | 1.039 |
| Radius ($R_\oplus$) | 1.045 |
| Orbital period (days) | 9.2075 |
| Transit duration (hr) | 1.05 |
| Inclination (deg) | 89.74 |
| Irradiation ($S_\oplus$) | 0.349 |
| Density (g cm$^{-3}$) | 5.042 |
| Surface gravity (m s$^{-2}$) | 9.37 |

### 2.3. LIFESIM

LIFE (Kammerer & Quanz 2018; Quanz et al. 2022) is a mission concept for a large, space-based, formation-flying, MIR, nulling interferometer observatory for the direct detection and atmospheric characterization of a large sample of terrestrial, potentially habitable exoplanets. ESA's Voyage 2050 Senior Committee report[11] recommended "the characterization of the atmosphere of temperate exoplanets in the mid-infrared" with "highest scientific priority" as a candidate topic for a future L-class mission in the ESA Science Programme. LIFE directly addresses this scientific theme. Typical observational cases for LIFE will be planets in the habitable zone of M stars at ~5 pc, and "Sun-like" FGK stars at ~10 pc (Dannert et al. 2022; Kammerer et al. 2022).

We use the LIFESIM software designed for LIFE to calculate the resulting detectability of features present in our simulated SMART-generated emission spectra using the same methods as described in Angerhausen et al. (2023). The current version of LIFESIM includes astrophysical noise sources, including stellar leakage and thermal emission from local zodiacal and exozodiacal dust. The software is designed to be flexible, allowing for the incorporation of instrumental noise terms in future iterations. An important feature of LIFESIM is its ability to provide a user-friendly means of predicting the expected S/N for future LIFE observations by considering various instrument dramatically beyond 5 μm for the NIRSpec prism detector. Any absorption features beyond that value would therefore be more detectable with MIRI LRS.

### 2.2. SMART

To generate synthetic emission spectra, we use the Spectra Mapping Atmospheric Radiative Transfer (SMART) model (Meadows & Crisp 1996; Crisp 1997). SMART is a 1D line-by-line, multistream, multiple-scattering radiative transfer model that is well validated by observations of Earth and solar system bodies (Tinetti et al. 2005; Robinson et al. 2014a, 2014b; Arney et al. 2014) and is often used to simulate the

---
[11] http://www.cosmos.esa.int/web/voyage-2050

**Table 2**
Overview of the Simulation Parameters Used in LIFESIM

| Parameter | Value |
| --- | --- |
| Quantum efficiency | 0.7 |
| Throughput | 0.05 |
| Minimum wavelength | 4 $\mu$m |
| Maximum wavelength | 18.5 $\mu$m |
| Spectral resolution | 50 |
| Interferometric baseline | 10–100 m |
| Apertures diameter | 2 m |
| Exo-zodiacal dust | 3 × local zodiacal dust |

**Note.** These are the same LIFE baseline values as those used in Quanz et al. (2022) or Angerhausen et al. (2023).

and target parameters. Further details can be found in Dannert et al. (2022).

The LIFESIM configuration for the presented output spectra followed the current LIFE "baseline" setup, featuring four apertures of 2 m diameter each, a broadband wavelength range spanning 4–18.5 $\mu$m, a throughput of 5%, and a spectral resolution set at $R = 50$. During the simulation, we assume an exo-zodiacal level equivalent to 3 times the local zodiacal density for the characterization observations. This assumption aligns with the findings of the HOSTS survey, which predicted the expected median emission level (as detailed in Ertel et al. 2020). The nulling baseline setup ranged from 10 to 100 m (this assumes prior knowledge of the planets through other surveys or detection in the LIFE detection phase, allowing for baseline optimization in each case). An overview of the LIFESIM simulation parameters is given in Table 2. For each case, we assumed the planet in the outer continuous habitable zone, specifically, at an orbital distance for which the solar constant is 0.53.

As in Angerhausen et al. (2024), we give two metrics for detectability: maximum difference in one band in units of sigma and the band-integrated S/N defined as:

$$(S/N)_{band} = \sqrt{\sum_{i=1}^{n}\left(\frac{\Delta y_i}{\sigma(y_i)}\right)^2}, \quad (1)$$

where $\Delta y_i$ is the difference between the spectrum containing the technosignature feature and a spectrum that does not in each of the $n$ spectral bins and $\sigma(y_i)$ is the LIFE sensitivity in the respective bin. More detailed studies would also include retrievals (Alei et al. 2022; Konrad et al. 2022), but we have shown that the qualitative detectability criteria used here translate well for quantitative retrievals (Angerhausen et al. 2024).

## 3. Synthetic Planetary Spectra Results

Here we describe the resulting synthetic planetary emission spectra for the artificial greenhouse gas scenarios described in Section 2. Where applicable, we quantified the relative detectability of these gas features with JWST and LIFE.

### 3.1. Transmission Spectra in the Mid-infrared

Figure 2 shows the resulting MIR (5–12 $\mu$m) transmission spectra of a hypothetical TRAPPIST-1 f and the various concentrations of the technosignature gases listed above. This wavelength region lies within the range of MIRI's LRS. We draw attention to the fact that even at 1 ppm, every technosignature gas produces maximum transmission depths that rival or exceed those of the 9.65 $\mu$m $O_3$ band, which is included within the plotted range. The PFC gases ($CF_4$, $C_2F_6$, and $C_3F_8$) contribute to discrete absorption maxima at several wavelengths. The strongest feature of $C_4$ is located at ~7.9 $\mu$m, with additional significant features at 5.25, 5.9, 6.5, and 9.2 $\mu$m. Each of these features spans $\gtrsim 0.02$ $\mu$m, depending on concentration. In contrast, $C_2F_6$ and $C_3F_8$ have more numerous features throughout the MIR that blend together at high concentration (potentially as a result of uncertainties in the input opacities, see the Discussion section), and much stronger absorption at wavelengths 8–9 $\mu$m. $C_3F_8$ additionally possesses a strong feature at ~9.9 $\mu$m. The strongest feature of $SF_6$ lies at ~10.7 $\mu$m with weaker features at 5.8, 6.3, 6.9, 7.2, 8.0, 8.8, and 11.4 $\mu$m. A combination of $C_3F_8$, $C_2F_6$, and $SF_6$ is thus strongly absorptive throughout the entire MIR region shown, particularly at higher greenhouse gas concentrations. We also examine $NF_3$, which has its strongest feature located at 11 $\mu$m, which is substantial at even 1 ppm. It also has weaker features at 5.2, 5.6, 6.5, 7.2, 8.8, and 9.8 $\mu$m. Simulated data points for the 100 ppm technosignatures gases, using 10 transits at $R = 50$, show a clear deviation from the black MIR spectrum with no technosignature gases.

In Table 3, we show the number of transits needed to detect each technosignature gas and gas combination at 5$\sigma$ on each simulated TRAPPIST-1 f with MIRI LRS according to the procedure described in Section 2.1. The combination of $C_2F_6$, $C_3F_8$, and $SF_6$ can be detected in five transits at 100 ppm and 10 transits at 10 ppm. In each scenario, the number of transits required to detect any of these technosignature gases at $\geqslant 1$ ppm concentration is orders of magnitude lower than for $O_3$ or $CO_2$, which are effectively undetectable with MIRI. We note that while uniquely retrieving the identities of these terraforming gases would likely require further characterization (see the Discussion), these results illustrate that unexpected MIR opacities consistent with a terraformed atmosphere, as simulated here, and are within the observational capabilities of JWST.

### 3.2. Transmission Spectra in the Near-infrared

Figure 3 shows simulated NIR (1.5–5 $\mu$m) spectra of TRAPPIST-1 f with 1, 10, and 100 ppm of $C_2F_6$, $C_3F_8$, $SF_6$, and an equal combination of the three. Notably, $C_2F_6$ and $C_3F_8$ produce substantial absorption features comparable to or greater than $CO_2$ at concentrations $\geqslant 10$ ppm. These two gases produce substantial absorption features in the 4–5 $\mu$m region that overlap with the 4.3 $\mu$m $CO_2$ band and the 4.8 $\mu$m $O_3$ band, but are much broader. In addition, a weaker band near 2.8 $\mu$m is apparent at high concentrations (~100 ppm) that overlaps with a separate $CO_2$ band. In contrast, the NIR features caused by $SF_6$ are muted and only apparent at concentrations $\gtrsim 10$ ppm. Even at 100 ppm concentrations of $SF_6$, its NIR spectral features are extremely limited when compared to those of $CO_2$. The combined NIR spectra are dominated by features from $C_2F_6$ and $C_3F_8$. Simulated data points for the 100 ppm technosignatures gases, using 10 transits at $R = 10$, again show a clear deviation from the black NIR spectrum with no technosignature gases.

We calculate the single transit S/N and the number of transits needed to detect these gases with JWST NIRSpec at 5$\sigma$

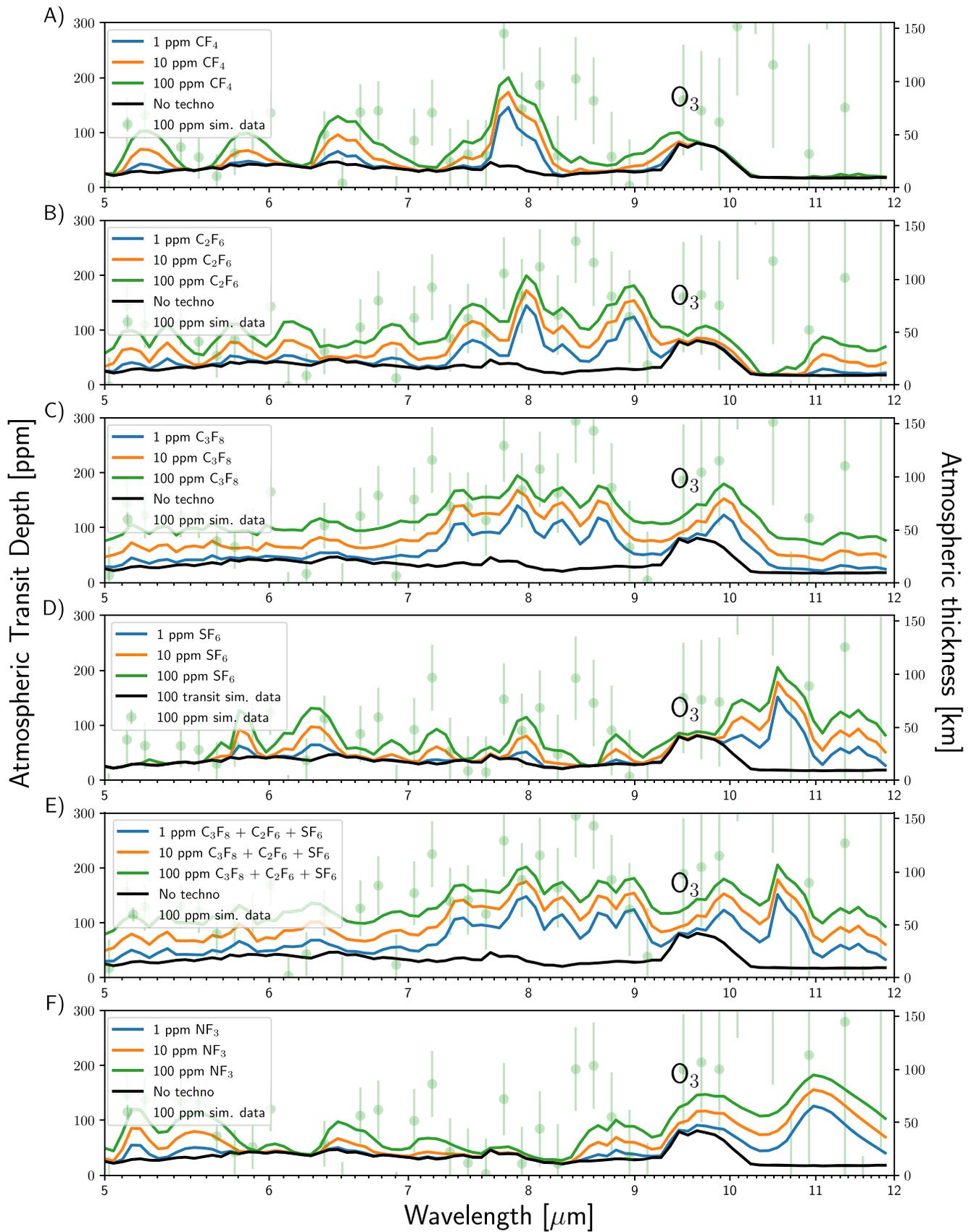

**Figure 2.** PSG-simulated MIR (5–12 μm) transmission spectra of an Earth-like TRAPPIST-1 f with 1–100 ppm of technosignature gases $CF_4$ (A), $C_2F_6$ (B), $C_3F_8$ (C), $SF_6$ (D), a combination of the preceding three gases (E), and $NF_3$ (F). Simulated data points for the 100 ppm cases are shown in green, for 10 transits and at a resolving power of $R = 50$ for clarity.

**Table 3**
Number of Transits Required to Reach a 5σ Detection (5σ Transit) with MIRI LRS for a Given Modeled Atmosphere

| Atmosphere | 5σ Transit |
| --- | --- |
| 100 ppm | |
| $C_2F_6 + C_3F_8 + SF_6$ | 5 |
| $C_2F_6$ | 6 |
| $C_3F_8$ | 8 |
| $SF_6$ | 19 |
| $NF_3$ | 16 |
| 10 ppm | |
| $C_2F_6 + C_3F_8 + SF_6$ | 10 |
| $C_2F_6$ | 13 |
| $C_3F_8$ | 20 |
| $SF_6$ | 57 |
| $NF_3$ | 52 |
| 1 ppm | |
| $C_2F_6 + C_3F_8 + SF_6$ | 25 |
| $C_2F_6$ | 38 |
| $C_3F_8$ | 60 |
| $SF_6$ | >100 |
| $NF_3$ | >100 |
| Molecular Comparison | |
| Earth-like $O_3$ | >100 |
| 378 ppm $CO_2$ | >100 |

in Table 4, again following the procedure described in Section 2.1. We find that the combination of all three gases is potentially detectable in as few as four transits if all are present at 100 ppm concentrations and as few as 14 transits at 10 ppm. Individually, each technosignature gas is less detectable, with $SF_6$ alone being functionally undetectable in the NIR with NIRSpec at any concentration ⩽ 100 ppm. However, at 100 ppm, $C_2F_6$ and $C_3F_8$ could be detected in around ∼40 transits, which is similar to the number of transits needed to detect 100 ppm of $CO_2$. Overall, we find a similar number of transits are required to detect these artificial greenhouse gases at high concentrations (∼100 ppm) in both the NIR and MIR, but there is an advantage to MIR observations in detecting intermediate concentrations (∼1–10 ppm). It is likely that access to both wavelength regimes would aid in full retrievals.

### 3.3. Emission Spectra in the Mid-infrared

We show the simulated emission spectra of Earth-twin planets with artificial greenhouse gases $CF_4$, $C_2F_6$, $C_3F_8$, $SF_6$, and $NF_3$ in Figure 4 assuming concentrations of 1, 10, and 100 ppm as we did for the transit transmission cases presented in Section 3.1. We further break out the scenarios with combinations of 1, 10, and 100 ppm each of $C_3F_8 + C_2F_6 + SF_6$ in Figure 5. The strongest apparent gases in emission will differ from those in transmission because the absorption depths depend not only on the concentration of the absorbing gas, but also the surface temperature of the planet, the temperature structure of the planet's atmosphere, and the interference with other absorbing gases that manifest differently in emission versus transit transmission (e.g., $H_2O$). We note here that even at 1 ppm, several of these technosignature gases, including $C_2F_6$, $C_3F_8$, and $SF_6$, display strong MIR absorption features that equal or exceed that of $O_3$ at 9.65 μm.

$CF_4$ has relatively weak features at 1 ppm, but substantial absorption at 7.9 and 9.2 μm at concentrations ⩾ 10 ppm. As in transmission, $C_2F_6$ produces strong features from 8 to 9 μm. At 100 ppm, $C_2F_6$ produces an additional strong feature near 11 μm. $C_3F_8$ has strong features throughout the 8–12 μm window region that are particularly strong from 8 to 10 μm. Furthermore, as seen in transit, $SF_6$ features are most prominent from 10 to 12 μm, with one strong feature at 8.8 μm apparent at 100 ppm. $NF_3$ has a strong feature at 11 μm apparent even at 1 ppm, with broad absorption from 10 to 12 μm at 10 ppm, and throughout the entire 8–12 μm window at concentrations of 100 ppm. At 100 ppm, some gases, such as $CF_4$ and $SF_6$, can show emission features because high optical depths are reached in the warm stratosphere. We note that the temperature structure is based on modern Earth, as described in Section 2. We describe the possible benefit of self-consistent calculations, and why they may not yet be possible, in Section 4. We quantify the detectability of these artificial greenhouse gas signatures with a hypothetical MIR direct-imaging observatory in the subsection below.

#### 3.3.1. Mid-infrared Observations with a LIFE-like Observatory

Tables 5 and 6 show the results for simulated LIFE observations with various combinations of host stars, technosignature gas combinations, stellar distances, and observation times. Figure 6 shows some exemplary cases including different stellar host star types (M8V, K6V, and G2V), distances (5–10 pc), and integration times (10–50 days). Maximum channel and band-integrated S/Ns are calculated according to the methodology given in Section 2.3. Virtually all cases seem to be observable (i.e., spectral features are detectable at or above the 5σ level) at typical observation times between 10 and 50 days, much less time than needed for the detection of typical biosignature gases such as $O_3$ at 9.65 μm (Alei et al. 2022; Konrad et al. 2022; Angerhausen et al. 2023, 2024). First hints at the presence of additional, unusual MIR absorbers would probably already be found after relatively short preliminary observations. These results indicate that an MIR instrument is the ideal choice to observe civilizations that manipulate their climates and that tests for these technosignature gases are essentially "free" since evidence for these features would be found even before typical biosignature gases become detectable. We provide an extended table of calculations for the detectability of 1, 10, and 100 ppm of $SF_6$, $C_2F_6$, $C_3F_8$, $NF_3$, and $CF_4$ applied to each host star (G2V, K6V, and M8V), distance (5 and 10 pc), and integration time (10 and 50 days) in Appendix B, Table 6.

### 4. Discussion

#### 4.1. Detectability of Artificial Greenhouse Gases

We have shown here that synthetic molecules previously proposed as potential terraforming gases on Mars (e.g., Marinova et al. 2005; Dicaire et al. 2013) can be detected on terrestrial exoplanets with JWST or a large, space-based MIR direct-imaging interferometer such as the LIFE concept (e.g., Kammerer & Quanz 2018) at the reasonable abundances necessary to impart a substantial climatic effect (i.e., ≳1 ppm in a 1 bar atmosphere). Specifically, we have shown that $CF_4$, $C_2F_6$, $C_3F_8$, $SF_6$, and $NF_3$, alone and in combination, can produce MIR (5–12 $\mu$m) transit signatures comparable to or greater than the 9.65 $\mu$m $O_3$ band at concentrations $\gtrsim 1$ ppm (Figure 2). We calculated the number of transits required to detect $C_2F_6$, $C_3F_8$, $SF_6$, $NF_3$, and equal combinations of the first three gases at 1, 10, and 100 ppm on TRAPPIST-1 f with MIRI LRS and found surprisingly few transits are required, as few as five for a $5\sigma$ detection for a combination of each gas at 100 ppm, and 10 transits for the same at 10 ppm (Table 3). Of the gases individually, $C_2F_6$ is the most detectable at any given concentration, while $C_3F_8$ is still strongly detectable, and $SF_6$ requires the most number of transits to be detected at any given concentration. $NF_3$ requires slightly fewer transits to detect than $SF_6$ (e.g., 16 versus 19 at 100 ppm). Nonetheless, $SF_6$ and $NF_3$ are still much more detectable than $O_3$ or $CO_2$ in an Earth-like atmosphere at concentrations > 1 ppm.

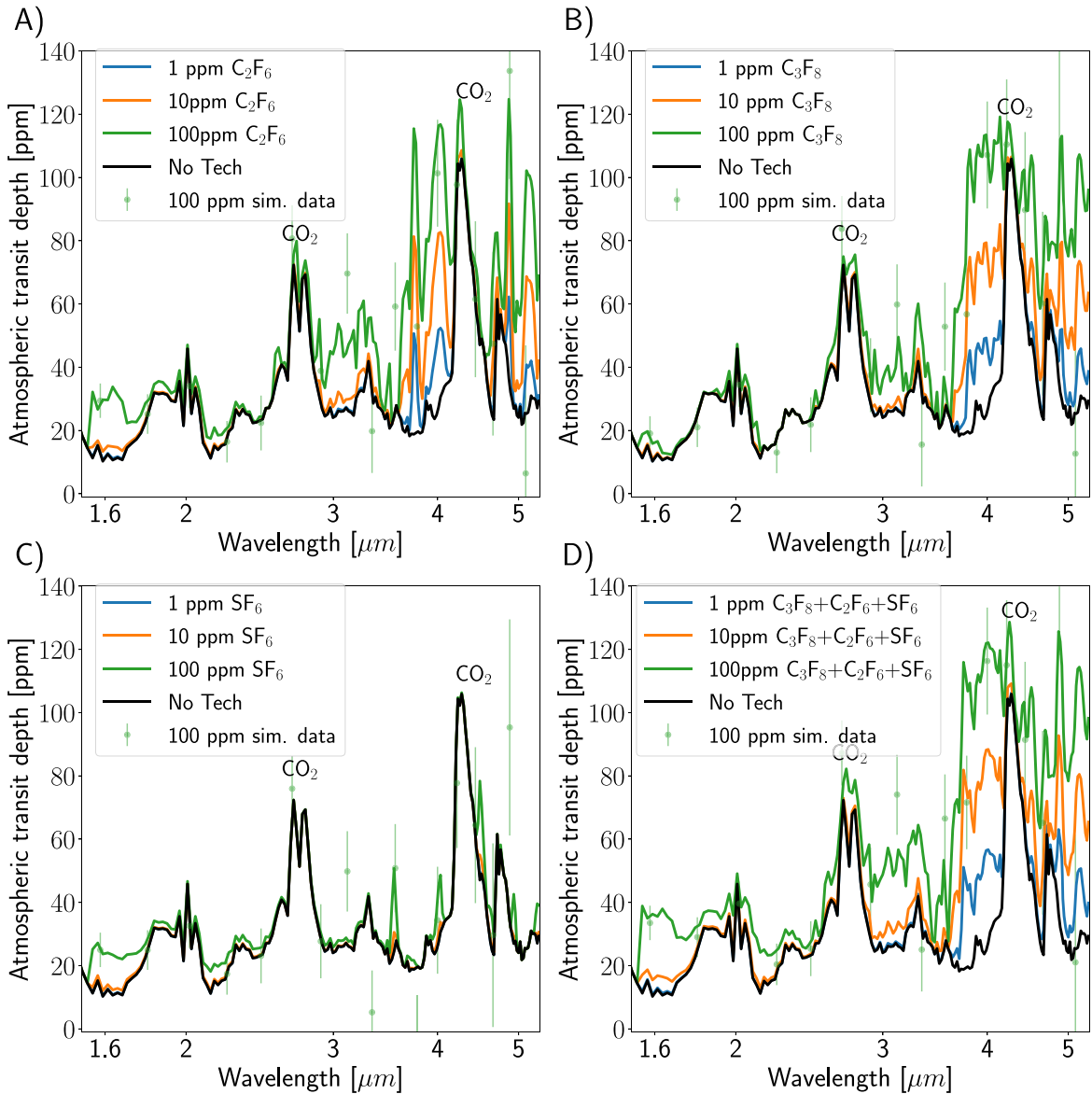

**Figure 3.** PSG-simulated NIR (1.5–5 $\mu$m) transmission spectra of the modeled TRAPPIST-1 f atmospheres with 1, 10, and 100 ppm of technosignature gases $C_2F_6$ (top left), $C_3F_8$ (top right), $SF_6$ (bottom left), and a combination of the preceding three gases (bottom right). Simulated data points for the 100 ppm cases are shown in green, for 10 transits and at a resolving power of $R = 10$ for clarity.

We repeated the above analysis in the NIR (1.5–5 $\mu$m) and calculated the number of transits required to detect each of the three gases ($C_2F_6$, $C_3F_8$, and $SF_6$) alone or in combination on TRAPPIST-1 f with JWST NIRSpec (Figure 3 and Table 4). We found that at concentrations of 100 ppm, a combination of all three gases can be detected in as few as four transits, and at 10 ppm in as few as 14 transits. However, each gas individually is far less detectable than the corresponding MIR case. While $C_2F_6$ and $C_3F_8$ produce spectral signatures comparable to $CO_2$ at 100 ppm, $SF_6$ alone is likely not reasonably detectable at concentrations $\leqslant 100$ ppm. Both our mid and NIR analyses of the detectability of these technosignature gases in transit compare favorably with past predictions for biosignature detectability on the TRAPPIST-1 planets. For example, multiple studies have found that the $CH_4$–$CO_2$ disequilibrium pair could be detected on an Archean-Earth-like TRAPPIST-1 e in ~10 transits (Krissansen-Totton et al. 2018a; Mikal-Evans 2021; Meadows et al. 2023), while we find that potentially even fewer transits would be required to detect $C_2F_6$ or $C_3F_8$ at 100 ppm, or in combination ($C_2F_6 + C_3F_8 + SF_6$) at 10 ppm. Moreover, these aforementioned studies find that detecting biogenic $O_2$ and $O_3$ may not be possible on the TRAPPIST-1 planets, while we find a range of scenarios that would allow detection of artificial greenhouse technosignature gases at concentrations of ~1 ppm.

**Table 4**
Number of Transits Required to Reach a 5σ Detection (5σ Transit) with NIRSpec for a Given Modeled Atmosphere

| Atmosphere | 5σ Transit |
|---|---|
| 100 ppm | |
| $C_2F_6 + C_3F_8 + SF_6$ | 4 |
| $C_2F_6$ | 36 |
| $C_3F_8$ | 41 |
| $SF_6$ | >100 |
| 100 ppm $CO_2$ | 48 |
| 378 ppm $CO_2$ | 25 |
| 10 ppm | |
| $C_2F_6 + C_3F_8 + SF_6$ | 14 |
| $C_2F_6$ | >100 |
| $C_3F_8$ | 83 |
| $SF_6$ | >100 |
| 1 ppm | |
| $C_2F_6 + C_3F_8 + SF_6$ | 80 |
| $C_2F_6$ | >100 |
| $C_3F_8$ | >100 |
| $SF_6$ | >100 |

We also calculated the MIR emitted light spectra for an Earth-twin planet with 1, 10, and 100 ppm of $CF_4$, $C_2F_6$, $C_3F_8$, $SF_6$, and $NF_3$ (Figure 4) and the corresponding detectability of $C_2F_6$, $C_3F_8$, and $SF_6$ with the LIFE concept mission (Figure 6 and Table 5). We find that in every case, the band-integrated S/Ns were >5σ for outer habitable zone Earths orbiting G2V, K6V, or TRAPPIST-1-like (M8V) stars at 5 and 10 pc and with integration times of 10 and 50 days. Importantly, the threshold for detecting these technosignature molecules with LIFE is more favorable than standard biosignatures such as $O_3$ and $CH_4$ at modern Earth concentrations, which can be accurately retrieved (Alei et al. 2022; Angerhausen et al. 2024), indicating meaningfully terraformed atmospheres could be identified through standard biosignatures searches with no additional overhead.

### 4.2. Fingerprinting Artificial Greenhouse through Infrared Anomalies

In principle, the abundances of technosignature molecules can be retrieved from transit transmission or MIR direct-imaging spectroscopy. For example, Lustig-Yaeger et al. (2023b) retrieve CFC-11 ($CCl_3F$) and CFC-12 ($CCl_2F_2$) abundances from reconstructed Earth transit spectra from the Canadian low-Earth orbit satellite SCISAT (Macdonald & Cowan 2019). However, as shown here, high concentrations of artificial greenhouse gases can act to fill the entire MIR window, blending features and potentially interfering with our ability to uniquely identify absorbing gas species (though a full exploration of every possible permutation is beyond the scope of this work). Indeed, an extraterrestrial civilization may be motivated to precisely optimize and calibrate a combination of artificial greenhouse gases to fill the MIR window and maximize warming potential at the lowest cost (see the related discussion as applied to terraforming Mars in, e.g., Marinova et al. 2005). In such a scenario, it may be difficult to uniquely identify and constrain the abundance of each contributing gas. However, in any case, meaningful radiative forcing necessitates high infrared opacity in the window region, which would be apparent in transit or emission spectroscopy. We propose that anomalously large MIR transit signatures or anomalously low infrared emission could serve as "first-pass" indicators of terraformed planets (see the concept illustrations in Figure 7). For transmission observations, the artificial nature of this anomalous absorption could be confirmed via higher-S/N MIR observations that would reveal individual molecular features, or diagnostic NIR features that would not be optimized for climatic impact. In emission spectra, higher-S/N observations may reveal features within the window, particularly at the wings of the strongest absorption features, which may show higher brightness temperatures than near the center.

MIR searches are important complements to biosignature searches in the UV/visible/NIR range because some gaseous species such as $CH_4$ and $N_2O$ (and $O_3$ if we neglect the UV) are more detectable via MIR observations at Earth-like concentrations (e.g., Alei et al. 2022; Konrad et al. 2022; Schwieterman et al. 2022; Angerhausen et al. 2024). Moreover, due to inner-working angle constraints, an NIR/visible/UV surveyor, such as the proposed Habitable Worlds Observatory (HWO), is most capable of characterizing habitable zone planets around Sun-like stars (LUVOIR-Team 2019). While the proposed LIFE mission is similarly well suited to detect and characterize "Earth twins" in synergy with HWO, it also has a unique discovery space in the habitable zones of late type stars [i.e., K and M versus G and F] due to its long baselines, which translate into very high spatial resolution (see, e.g., Carrión-González et al. 2023). Given the impetus to characterize temperate rocky exoplanets in the MIR to search for signs of habitability and life, a commensal search for artificial greenhouse gases on the same targets would not impose further costs (NASA Technosignatures Workshop Participants 2018; Haqq-Misra et al. 2022b), and could reveal these signatures if their absorption features are comparable to the habitability markers that are sought by biosignature searches. We quantitatively demonstrate that these features are indeed comparable or exceed those of standard biosignature gases at abundances ≧ 1 ppm.

### 4.3. The Case for Long-lived, Low-toxicity Greenhouses Gases as Compelling Planetary Technosignatures

As mentioned in the Introduction, a civilization may be motivated to modify the climate of a planet to forestall imminent global cooling on their home world, or to terraform an otherwise uninhabitable planet within their system (or beyond), as humans have proposed to do with Mars. Alternatively, a civilization may seek to reduce the amount of $CO_2$ necessary to warm an already habitable planet in the outer habitable zone, which could require several bars of $CO_2$ to remain clement (Kopparapu et al. 2013; Shields et al. 2016; Schwieterman et al. 2019). For example, Fauchez et al. (2019) find that 1–10 bars of $CO_2$ is required to warm the surface of TRAPPIST-1 f or g. These levels of $CO_2$ impose substantial physiological challenges on complex aerobic organisms (e.g., Schwieterman et al. 2019; Ramirez 2020), so while these environments may be "habitable" to microbes, it does not necessarily follow that they would be suitable for higher forms of organic life. The radiative forcing otherwise provided by massive amounts of $CO_2$ could instead be provided by alternative molecules that more effectively absorb within the MIR window (8–12 $\mu$m), a wavelength range that is predetermined by the blackbody peak of a habitable planetary surface (e.g., ~300 K).

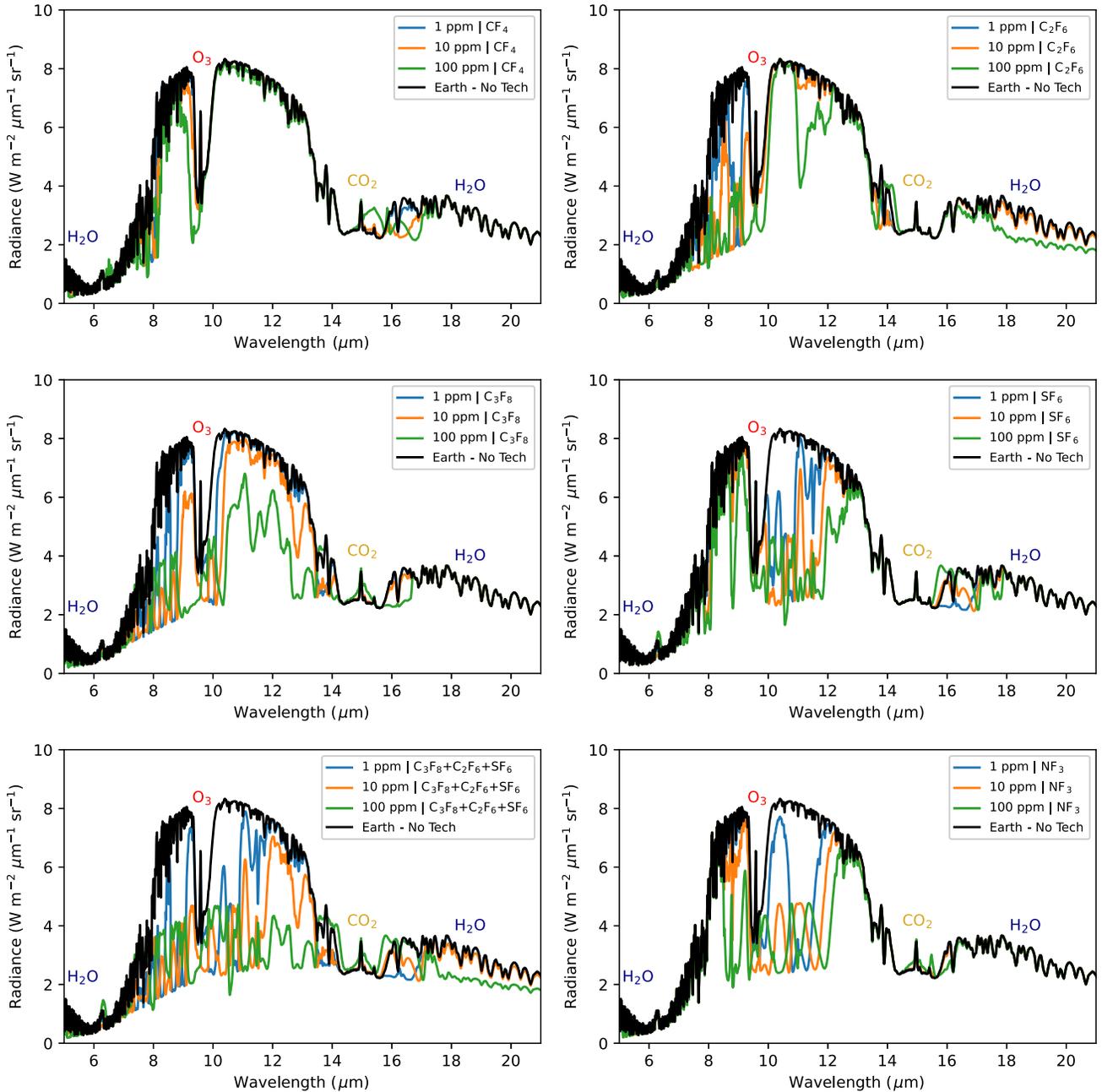

**Figure 4.** SMART-simulated MIR (5–21 $\mu$m) emission spectral radiances of an Earth-twin planet with 1–100 ppm of technosignature gases $CF_4$ (top left), $C_2F_6$ (top right), $C_3F_8$ (middle left), $SF_6$ (middle right), a combination of the preceding three gases (bottom left), and $NF_3$ (bottom right).

Fluorine-bearing gases including PFCs (e.g., $CF_4$, $C_2F_6$, and $C_3F_8$), $SF_6$, and $NF_3$ have been proposed as artificial greenhouse gases for terraforming Mars for multiple reasons, including their high MIR opacity, long atmospheric lifetimes, reduced reactivity with atmospheric $O_3$, and low toxicity to life due to their chemical inertness (Marinova et al. 2005). On a molecule-per-molecule basis, the radiative forcing of the artificial greenhouse gases explored here are orders of magnitude greater than $CO_2$, and so bars of $CO_2$ could hypothetically be replaced by a much smaller atmospheric loading of these technosignature gases (Marinova et al. 2005). For example, the global warming potential (GWP) of $C_2F_6$ is on the order of 10,000 times that of $CO_2$ on a 100 yr timescale (see Table 1 in Mühle et al. 2010). $SF_6$ has a 100 yr GWP of 23,500 (Myhre et al. 2013).

These gases are primarily produced on Earth today in small quantities as a by-product of industrial processes. $SF_6$ is manufactured as an insulator and has a modern atmospheric concentration of 11 ppt (Lan et al. 2023). It has a lifetime of approximately ~1000 yr, though estimates vary by a few hundred years, and is primarily destroyed in the ionosphere via collisions with energetic electrons (Kovács et al. 2017; Ray et al. 2017). PFCs are produced primarily as a by-product of aluminum smelting and semiconductor production and have atmospheric lifetimes of ~2000–50,000 yr (Mühle et al. 2010), and modern concentrations of 87 ppt ($CF_4$), 5 ppt ($C_2F_6$), and 0.7 ppt ($C_3F_8$; Prinn et al. 2018a, 2018b). PFCs are destroyed by molecule–ion reactions in the ionosphere (Lodders & Fegley 2015). $NF_3$ has been used as a replacement for PFCs in circuit etching and other chemical processes, but has an atmospheric lifetime of 1020 yr and a GWP of 7370 (Maione et al. 2013; Totterdill et al. 2016). It has a modern concentration of 2.5 ppt (Prinn et al. 2018a, 2018b). While the concentrations of these gases are small, their high GWP values make these gases important targets for emission reduction to arrest the worst impacts of global climate change (e.g., Illuzzi & Thewissen 2010).

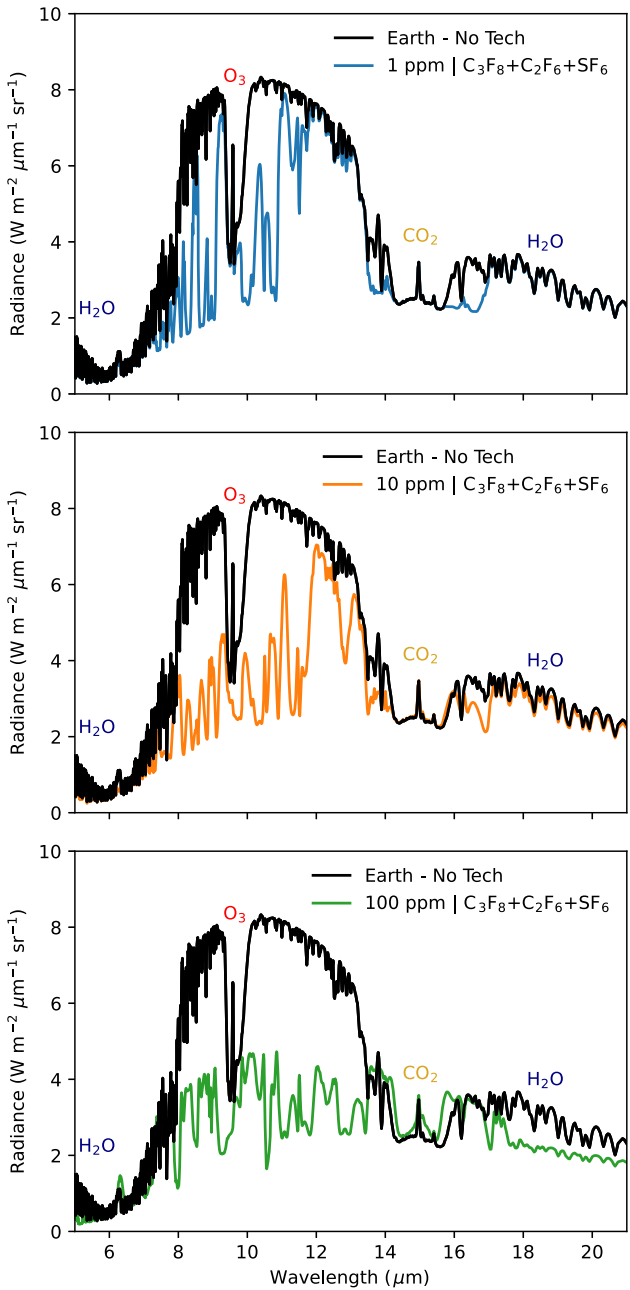

**Figure 5.** Similar to Figure 4, but breaking out SMART-simulated emission spectra with combinations of $C_3F_8 + C_2F_6 + SF_6$ each of which with concentrations of 1 ppm (top panel), 10 ppm (middle panel), and 100 ppm (bottom panel). The Earth without technosignature gases is shown for comparison in each panel (black line).

The vast majority of these gases present in the modern atmosphere have been produced via anthropogenic activity, though small abiotic sources exist. $CF_4$ and $SF_6$ are produced via the weathering of fluorite-bearing minerals and have measured preindustrial background concentrations of ~35 and ~0.01 ppt based on ice core samples (Harnisch & Eisenhauer 1998; Mühle et al. 2010). The preindustrial concentrations of $C_2F_6$, $C_3F_8$, and $NF_3$ are either small (<0.01 ppt) or nonexistent (Arnold et al. 2013; Trudinger et al. 2016). In addition to low-abiotic sources, Seager et al. (2023) argue that fully fluorinated $NF_3$ and $SF_6$ are unlikely to be made biologically (by nontechnological life) because of life's aversion to using fluorine except in limited circumstances and is not known to produce fully fluorinated volatile gases. See also a related discussion of the likelihood of fluorine-bearing gaseous biosignatures in Section 5.2 of Leung et al. (2022).

While all technosignature scenarios are speculative, we argue that it is unlikely fluorine-bearing technosignature gases will accumulate to detectable levels in a technosphere due only to inadvertent emission of industrial pollutants (or volcanic production), and that our best opportunity to detect them likely requires their use at higher abundances for the intentional modification of a planet's climate. These gases exist at parts-per-trillion levels on Earth's atmosphere, which are undetectable over interstellar distances with near or intermediate future telescopes, and yet nonetheless meaningfully contribute to our civilizational challenge of global climate change because of their large GWP values. PFCs, $SF_6$, and $NF_3$ are regulated by the UN Framework Convention on Climate Change (UNFCCC) with reporting requirements for all member nations (UNFCCC COP 19, 2013). These concentrations are approximately 6 orders of magnitude smaller than those we examine in this work. However, as argued above and in our Introduction, purposeful climate modification (including terraforming of otherwise nonhabitable planets) provides a plausible motivation for a technological civilization to maintain large concentrations of artificial fluorine-bearing greenhouse gases over long timescales.

### 4.4. Limitations of Current Work and Need for Refined Opacities

We presented first-pass simulations of artificial greenhouse gases at prescribed mixing ratios (1, 10, and 100 ppm in a 1 bar atmosphere). We did not simulate self-consistent climates, which could be an important consideration for future work. Planetary climate will depend on numerous factors including instellation, surface albedo, cloud composition and cloud coverage, total atmospheric pressure, mixing ratios of nontechnosignature gases, surface water availability, and many more. It is beyond the scope of this work to analyze the effect of each of these variables on technosignature gas detectability, however, we will briefly note below future areas for improvement.

In our transmission spectra, our clouds are prescribed, rather than self-consistent, which could lead to differences in the lower atmosphere corresponding to smaller transit depths. However, because the transit altitudes for artificial greenhouse gases are far above the troposphere (~15 km), the impact on our results will be minimal. Fauchez et al. (2019) find that the transits needed to detect $CH_4$, $CO_2$, and $O_3$ on TRAPPIST-1 f

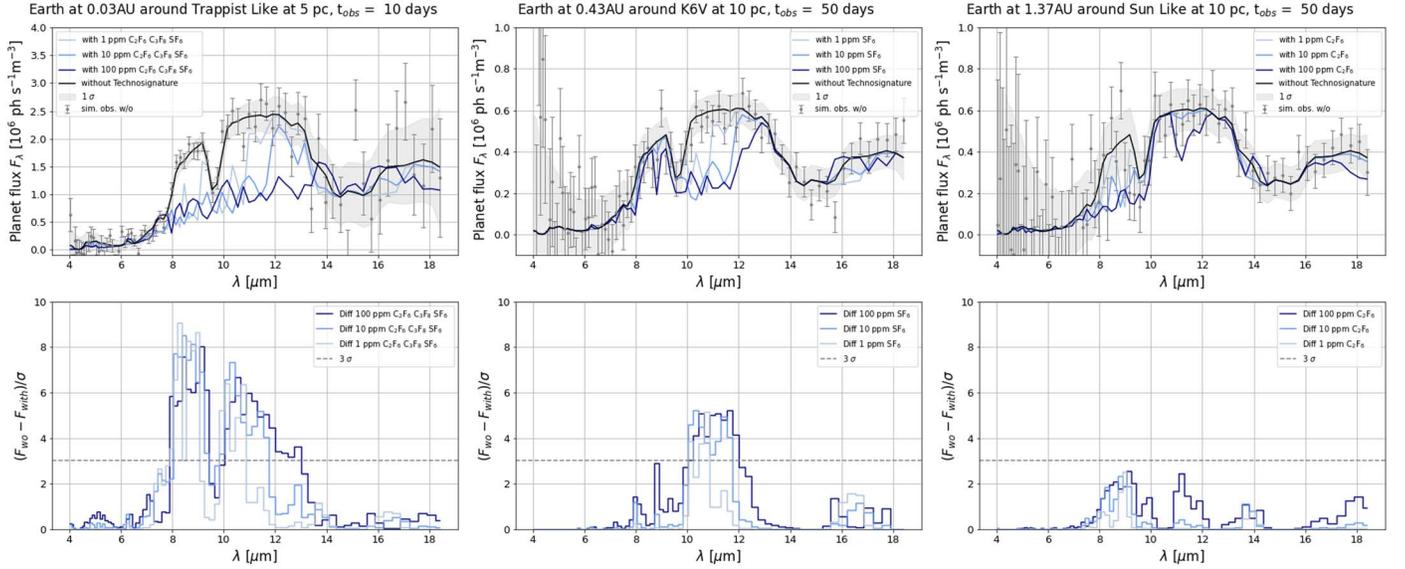

**Figure 6.** Three LIFE example artificial greenhouse gas detection cases (parameters in the titles). From left to right: least challenging (TRAPPIST-1-like host at 5 pc, 10 days, a combination of three technosignature gases), average (K6V host at 10 pc, 50 days, $SF_6$ only), and most challenging (Sun-like host at 10 pc, 50 days, $C_2F_6$ only). Top: planet flux for atmospheres with and without the feature. The gray area represents the $1\sigma$ sensitivity; the gray error bars show an individual simulated observation. Bottom: statistical significance of the detected differences between an atmospheric model with and without the technosignature gas(es).

**Table 5**
Detectability of the $C_2F_6 + C_3F_8 + SF_6$ Combination of Technosignature Gases with LIFE

| Host Star | Distance | $T_{obs}$ | Feature | Band Integrated ($\sigma$) | Max. Channel ($\sigma$) |
|---|---|---|---|---|---|
| Sun like | 5 pc | 10 days | 1 ppm all | 53.9 | 4.4 |
| Sun like | 5 pc | 10 days | 10 ppm all | 77.0 | 4.5 |
| Sun like | 5 pc | 10 days | 100 ppm all | 101.6 | 4.8 |
| Sun like | 10 pc | 50 days | 1 ppm all | 54.9 | 5.2 |
| Sun like | 10 pc | 50 days | 10 ppm all | 81.1 | 5.2 |
| Sun like | 10 pc | 50 days | 100 ppm all | 107.1 | 5.4 |
| K6V | 5 pc | 10 days | 1 ppm all | 101.7 | 10.4 |
| K6V | 5 pc | 10 days | 10 ppm all | 153.6 | 10.7 |
| K6V | 5 pc | 10 days | 100 ppm all | 206.3 | 11.2 |
| K6V | 10 pc | 50 days | 1 ppm all | 76.6 | 7.3 |
| K6V | 10 pc | 50 days | 10 ppm all | 112.8 | 7.4 |
| K6V | 10 pc | 50 days | 100 ppm all | 144.4 | 7.3 |
| Trappist like | 5 pc | 10 days | 1 ppm all | 138.3 | 12.7 |
| Trappist like | 5 pc | 10 days | 10 ppm all | 187.7 | 12.0 |
| Trappist like | 5 pc | 10 days | 100 ppm all | 207.8 | 11.3 |
| Trappist like | 10 pc | 50 days | 1 ppm all | 27.0 | 2.6 |
| Trappist like | 10 pc | 50 days | 10 ppm all | 38.5 | 2.4 |
| Trappist like | 10 pc | 50 days | 100 ppm all | 46.3 | 2.1 |

**Note.** Results for each technosignature gas individually at 1, 10, and 100 ppm abundance can be found in Appendix B.

are only different by ~20% between cloudy and cloud-free scenarios generated with a 3D global circulation model (GCM; see their Table 4). As we show, artificial greenhouse gases can produce features with substantially larger transit depths than these species at the concentrations we examined (1–100 ppm). In emitted light, we do not include clouds; however, we have previously found that clouds do not substantially impact retrieved abundances of key gases in an Earth-like atmosphere (e.g., Konrad et al. 2023). Assuming clouds have wavelength-independent effects on the emission spectra, they will decrease the emitted light by no more than a factor of 2, and thus the reported band-integrated S/Ns by no more than a factor of $\sqrt{2}$. This would not impact any of our conclusions. Future work should incorporate these greenhouse gases into 3D GCMs to self-consistently simulate climate and include realistic clouds.

Emitted light spectra are sensitive to the temperature profile of the planet's atmosphere. As a first-pass assumption, we assumed the modern Earth's temperature profile. A habitable surface temperature is consistent with the intent of terraforming. However, some greenhouse gases absorb shortwave (UV/visible/NIR) radiation, and can heat the upper atmosphere causing a temperature inversion, as $O_3$ does in Earth's atmosphere. Shortwave heating in the upper atmosphere would not only impact the upper atmosphere temperature structure, and thus the emitted light spectrum, but also generate an antigreenhouse effect that will partly act against the greenhouse effect of the gas of interest. We do not include this potential effect in our work. Indeed the opacity data for our all five gases examined here ($CF_4$, $C_2F_6$, $C_3F_8$, $SF_6$, and $NF_3$) extend to wavelengths no shorter than 1.5 $\mu$m (Sharpe et al. 2004), which makes such calculations currently impossible. Opacity measurements at the shortest NIR wavelengths (∼1.5–3 $\mu$m) are at or near the detection limit and also have substantial uncertainties. No climate calculations will be truly self-consistent without this shortwave opacity data.

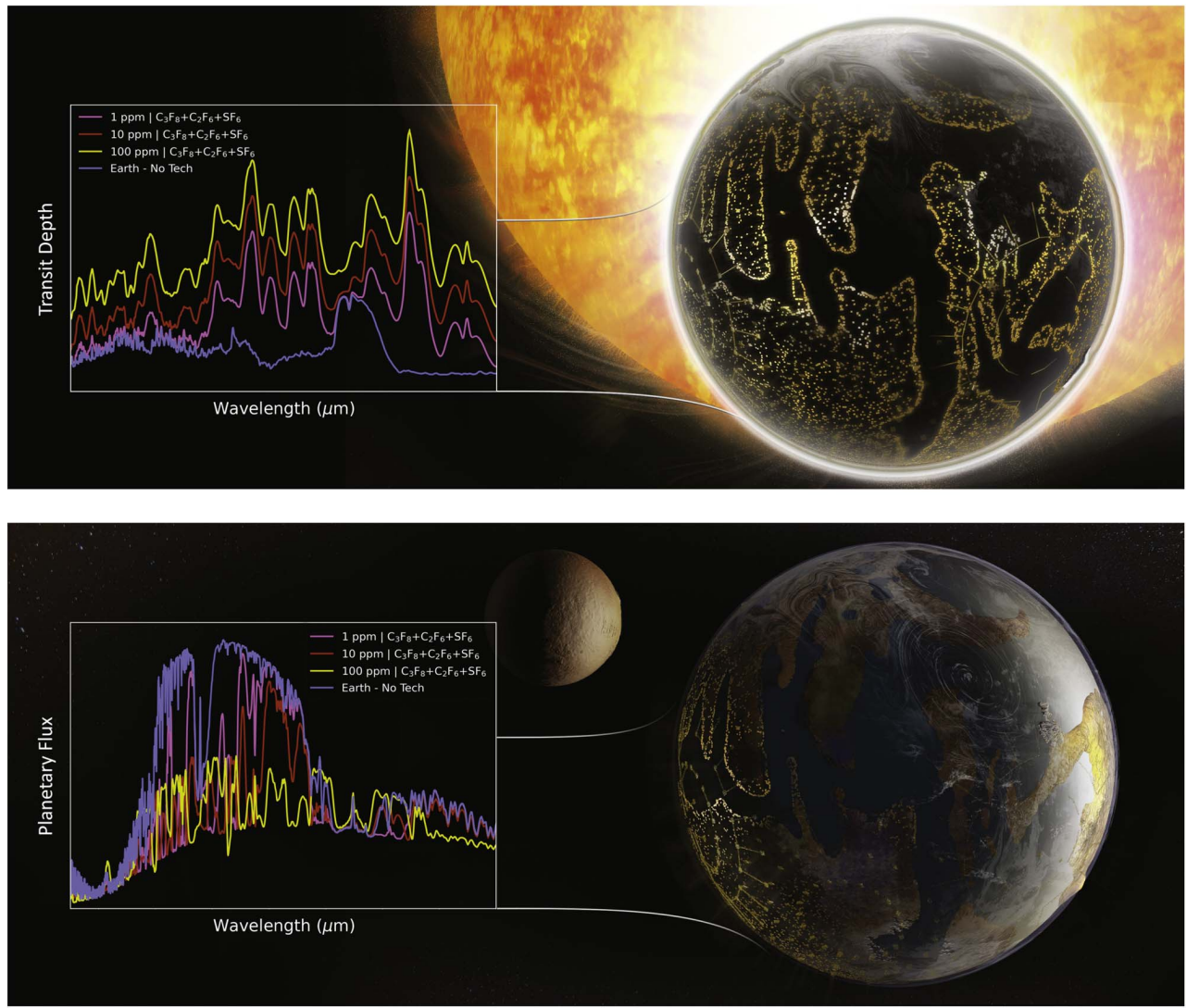

**Figure 7.** Concept figure illustrating a hypothetical Earth-like inhabited planet terraformed with various combined abundances of artificial greenhouse gases $C_3F_8$, $C_2F_6$, and $SF_6$ and its resulting qualitative MIR transmission (top) and emission (bottom) spectra. The figure conveys the anomalously high absorption opacity that may be detected when characterizing an exoplanet whose climate is modified by these artificial gases, which span the key MIR window wavelengths. (Graphic artist: Sohail Wasif, UC Riverside.)

Because we do not have robust opacities at wavelengths relevant to direct imaging of terrestrial planets in reflected light (∼0.2–2.0 $\mu$m), we cannot yet assess the potential detectability of these technosignature gases with the IR/visible/UV surveyor recommended by the 2020 Astronomy & Astrophysics Decadal Survey (National Academies of Sciences, Engineering, & Medicine & others 2023), currently named the HWO. We recommend future opacity measurements for these gases, particularly at visible and NIR wavelengths that are currently unavailable or have high uncertainties. Currently unavailable line-by-line data may be important for assessing the plausibility of detecting technosignature gases via high-resolution spectroscopy with extremely large telescopes (e.g., Currie et al. 2023; Hardegree-Ullman et al. 2023).

It is also important to mention that our analysis does not assume any noise floor and supposes that the star is perfectly homogeneous. Yet, unocculted starspots (spots on the star's surface not covered by the planet during its transit) cause stellar flux variability. Because starspots are cooler and darker than the surrounding stellar photosphere, they reduce the total stellar flux when present on the visible disk of the star. This variability can be misinterpreted as changes in the planet's transit depth, leading to incorrect interpretations of the planet's atmospheric properties. Specifically, the effect of starspots on observed flux is wavelength dependent, with a more significant impact at shorter wavelengths (Lim et al. 2023) due to the contrast between the cooler spots and the hotter photosphere. This spectral dependence can mimic the wavelength-dependent signals expected from atmospheric features, complicating the differentiation between true atmospheric signals and stellar activity effects. Lim et al. (2023), and references therein, highlight how the flux variability caused by unocculted starspots imposes an effective noise floor on transmission spectroscopy measurements, and therefore, a lower limit on the detectability of atmospheric features. For planets orbiting active stars like TRAPPIST-1, this noise floor can be particularly problematic because the small size of the star and the planets means that even small changes in stellar flux can significantly impact the measured transit depths. Any atmospheric spectral features with depths lower than 10 ppm, such as $SF_6$ in the NIR spectrum of Figure 3, are likely to be difficult to detect against this noise floor. The presence of starspots introduces a level of variability that can easily exceed 10 ppm, especially when considering the cumulative effect over multiple transits and the inherent variability in spot coverage over time. Detecting subtle atmospheric features, such as those indicative of certain gases or atmospheric compositions, becomes challenging under these conditions but should improve going to longer wavelengths (see Figure 8 in Appendix A), showcasing the utility of extending planetary characterization into the MIR.

## 5. Conclusions

We simulated the spectral impact of artificial greenhouse gases on the transmission and emission spectra of Earth-like exoplanets. Specifically, we examined the spectral impact of PFC species $CF_4$, $C_2F_6$, and $C_3F_8$, in addition to $SF_6$ and $NF_3$ at concentrations of 1, 10, and 100 ppm. These concentrations are consistent with those required to substantially modify planetary climate. We further calculated the number of transits required to detect 1, 10, and 100 ppm of $C_2F_6$, $C_3F_8$, $SF_6$, and $NF_3$, or a combination of the first three listed gases on TRAPPIST-1 f with JWST MIR/LRS and NIRSpec. We find that a combination of 1[10](100) ppm each of $C_2F_6$, $C_3F_8$, and $SF_6$ can be detected with an S/N $\geqq 5$ in as few as 25[10](5) transits with MIRI/LRS. Each gas individually can be detected at concentrations of >1 ppm in under 100 transits (except for $SF_6$ and $NF_3$ at 1 ppm), and as few as six transits with 100 ppm ($C_2F_6$). With JWST NIRSpec we find that a combination of 1[10](100) ppm each of $C_2F_6$, $C_3F_8$, and $SF_6$ can be detected in as few as 80[14](4) transits. Individually, each gas requires at least 36 transits at 100 ppm ($C_2F_6$). Overall, in transit, $C_2F_6$ and $C_3F_8$ both produce stronger spectral signatures than $SF_6$ and $NF_3$ at equal concentrations. We analyze the detectability of $C_2F_6$, $C_3F_8$, and $SF_6$ individually and in combination with the MIR direct-imaging LIFE mission concept, and find that all three gases at concentrations $\geqslant 1$ ppm produce large spectral features that rival or exceed $O_3$ at 9.65 $\mu$m and have band-integrated S/N $> 5\sigma$ for all scenarios examined.

We propose that the observation of anomalous MIR (or NIR) absorption in an exoplanetary atmosphere would be consistent with the presence of artificial greenhouse gases in a candidate technosphere. Such anomalous absorption may even be the first indication of a technosphere before any distinguishing features of individual technosignature gases can be positively confirmed. But such a detection would also raise the possibility that this anomalous absorption is the result of as-yet-unknown planetary processes that do not occur on Earth. Resolving this ambiguity between a technosignature and false positives will ultimately require extended observations to fully characterize the exoplanetary environment.

Future work should examine climatically self-consistent scenarios, though we note that additional opacity information is likely required, especially at NIR and visible wavelengths. Short-wavelength opacity data of technosignature gases would also be required to assess their detectability with an IR/visible/UV surveyor mission such as HWO. We conclude that artificial greenhouse gases are viable technosignatures that can be detected during otherwise routine planetary characterization operations at infrared wavelengths. Both positive or negative results would meaningfully inform the search for life elsewhere.

## Acknowledgments

This study resulted in part from the TechnoClimes workshop (2020 August 3–7, http://technoclimes.org), which was supported by the NASA Exobiology program under award 80NSSC20K1109. We acknowledge additional support from the NASA Interdisciplinary Consortium for Astrobiology Research (ICAR) Program under grant Nos. 80NSSC21K0905, 80NSSC23K1399, and 80NSSC21K0594. J.H.M. and R.K.K. also acknowledge support from the NASA Exobiology program under grant 80NSSC22K1009. The statements in this study do not necessarily reflect the views of NASA or any other funding agency or institution. D.A.'s work has been carried out within the framework of the National Centre of Competence in Research PlanetS supported by the Swiss National Science Foundation under grants 51NF40_182901 and 51NF40_205606. We thank UCR graphic artist Sohail Wasif for his assistance in creating the concept figure shown in the Discussion section. We thank Kevin Stevenson for helpful input regarding the noise modeling. We are also grateful to the anonymous referee for comments that allowed us to improve the paper.

*Software:* SMART (Meadows & Crisp 1996; Crisp 1997), PSG (Villanueva et al. 2018, 2022), LIFESIM (Dannert et al. 2022), Matplotlib (Hunter 2007), and NumPy (Harris et al. 2020).

## Appendix A
## Transit Depth Uncertainty for NIRSpec Prism and MIRI LRS

Figure 8 shows the transit depth uncertainty (parts per million) corresponding to the simulations shown in Figures 2 and 3.

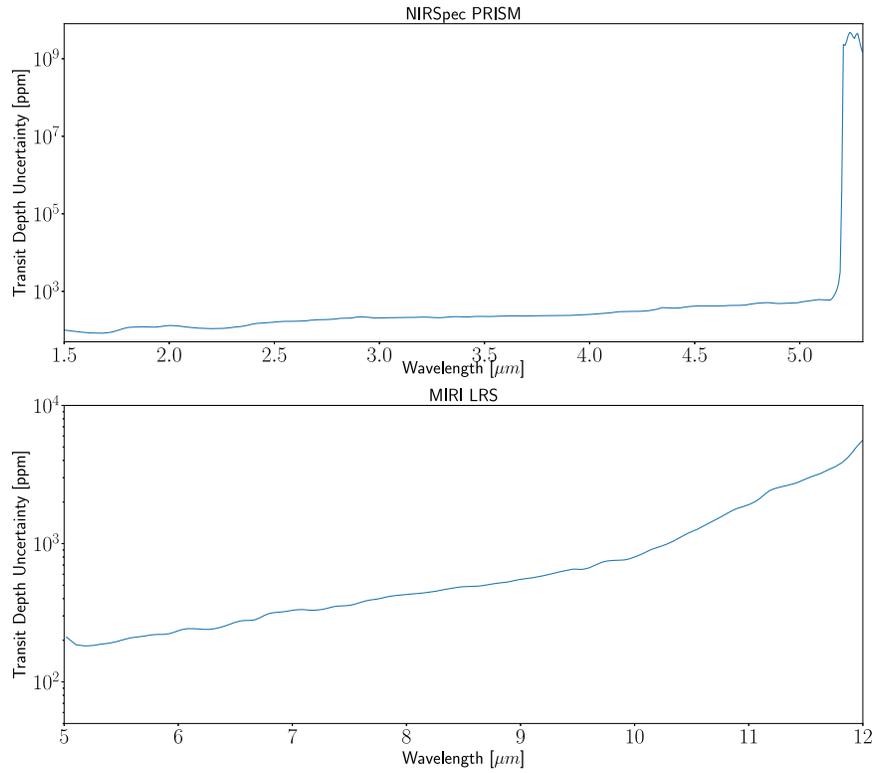

**Figure 8.** Top: transit depth uncertainty (in units of ppm) for the NIRSpec prism (restricted in the wavelength range where technosignature gas absorption are located). Bottom: transit depth uncertainty (in units of ppm) for MIRI LRS.

## Appendix B
## Additional Tables for the LIFE Simulations

Table 6 provides the band-integrated and maximum in channel S/Ns for individual technosignature gases in an Earth-like atmosphere calculated using LIFESIM as described in Sections 2.3 and 3.3.1. The corresponding emission spectra are calculated with SMART as described in Section 2.2 and shown in Section 3.3.

**Table 6**
Detectability of Individual Technosignature Gases with LIFE Calculated with LIFESIM

| Host Star | Distance | $T_{obs}$ | Feature | Band-int. ($\sigma$) | Max. Channel ($\sigma$) |
|---|---|---|---|---|---|
| $SF_6$ | | | | | |
| Sun like | 5 pc | 10 days | 1 ppm $SF_6$ | 25.3 | 4.4 |
| Sun like | 5 pc | 10 days | 10 ppm $SF_6$ | 41.7 | 4.5 |
| Sun like | 5 pc | 10 days | 100 ppm $SF_6$ | 63.6 | 4.6 |
| Sun like | 10 pc | 50 days | 1 ppm $SF_6$ | 27.1 | 5.2 |
| Sun like | 10 pc | 50 days | 10 ppm $SF_6$ | 46.2 | 5.2 |
| Sun like | 10 pc | 50 days | 100 ppm $SF_6$ | 69.5 | 5.5 |
| K6V | 5 pc | 10 days | 1 ppm $SF_6$ | 52.9 | 10.4 |
| K6V | 5 pc | 10 days | 10 ppm $SF_6$ | 91.0 | 10.5 |
| K6V | 5 pc | 10 days | 100 ppm $SF_6$ | 135.2 | 11.3 |
| K6V | 10 pc | 50 days | 1 ppm $SF_6$ | 36.1 | 7.3 |
| K6V | 10 pc | 50 days | 10 ppm $SF_6$ | 63.4 | 7.4 |
| K6V | 10 pc | 50 days | 100 ppm $SF_6$ | 95.3 | 7.3 |
| Trappist like | 5 pc | 10 days | 1 ppm $SF_6$ | 42.5 | 9.0 |
| Trappist like | 5 pc | 10 days | 10 ppm $SF_6$ | 83.1 | 10.3 |
| Trappist like | 5 pc | 10 days | 100 ppm $SF_6$ | 134.9 | 9.9 |
| Trappist like | 10 pc | 50 days | 1 ppm $SF_6$ | 7.4 | 1.5 |
| Trappist like | 10 pc | 50 days | 10 ppm $SF_6$ | 15.0 | 1.8 |
| Trappist like | 10 pc | 50 days | 100 ppm $SF_6$ | 25.6 | 1.9 |
| $C_2F_6$ | | | | | |
| Sun like | 5 pc | 10 days | 1 ppm $C_2F_6$ | 14.3 | 3.8 |
| Sun like | 5 pc | 10 days | 10 ppm $C_2F_6$ | 28.3 | 3.6 |

**Table 6**
(Continued)

| Host Star | Distance | $T_{obs}$ | Feature | Band-int. ($\sigma$) | Max. Channel ($\sigma$) |
|---|---|---|---|---|---|
| Sun like | 5 pc | 10 days | 100 ppm $C_2F_6$ | 56.3 | 3.8 |
| Sun like | 10 pc | 50 days | 1 ppm $C_2F_6$ | 13.6 | 3.5 |
| Sun like | 10 pc | 50 days | 10 ppm $C_2F_6$ | 27.1 | 3.3 |
| Sun like | 10 pc | 50 days | 100 ppm $C_2F_6$ | 55.1 | 3.6 |
| K6V | 5 pc | 10 days | 1 ppm $C_2F_6$ | 23.4 | 6.2 |
| K6V | 5 pc | 10 days | 10 ppm $C_2F_6$ | 47.1 | 5.6 |
| K6V | 5 pc | 10 days | 100 ppm $C_2F_6$ | 100.0 | 7.0 |
| K6V | 10 pc | 50 days | 1 ppm $C_2F_6$ | 19.9 | 5.3 |
| K6V | 10 pc | 50 days | 10 ppm $C_2F_6$ | 39.3 | 5.0 |
| K6V | 10 pc | 50 days | 100 ppm $C_2F_6$ | 76.3 | 5.4 |
| Trappist like | 5 pc | 10 days | 1 ppm $C_2F_6$ | 48.6 | 12.2 |
| Trappist like | 5 pc | 10 days | 10 ppm $C_2F_6$ | 94.4 | 12.3 |
| Trappist like | 5 pc | 10 days | 100 ppm $C_2F_6$ | 148.2 | 12.2 |
| Trappist like | 10 pc | 50 days | 1 ppm $C_2F_6$ | 10.0 | 2.3 |
| Trappist like | 10 pc | 50 days | 10 ppm $C_2F_6$ | 20.7 | 2.5 |
| Trappist like | 10 pc | 50 days | 100 ppm $C_2F_6$ | 34.1 | 2.4 |
| $C_3F_8$ | | | | | |
| Sun like | 5 pc | 10 days | 1 ppm $C_3F_8$ | 23.5 | 3.2 |
| Sun like | 5 pc | 10 days | 10 ppm $C_3F_8$ | 44.2 | 4.0 |
| Sun like | 5 pc | 10 days | 100 ppm $C_3F_8$ | 77.9 | 4.8 |
| Sun like | 10 pc | 50 days | 1 ppm $C_3F_8$ | 23.3 | 3.1 |
| Sun like | 10 pc | 50 days | 10 ppm $C_3F_8$ | 45.0 | 4.4 |
| Sun like | 10 pc | 50 days | 100 ppm $C_3F_8$ | 82.4 | 5.4 |
| K6V | 5 pc | 10 days | 1 ppm $C_3F_8$ | 40.9 | 6.0 |
| K6V | 5 pc | 10 days | 10 ppm $C_3F_8$ | 82.0 | 8.5 |
| K6V | 5 pc | 10 days | 100 ppm $C_3F_8$ | 157.0 | 11.0 |
| K6V | 10 pc | 50 days | 1 ppm $C_3F_8$ | 33.8 | 4.6 |
| K6V | 10 pc | 50 days | 10 ppm $C_3F_8$ | 63.9 | 6.3 |
| K6V | 10 pc | 50 days | 100 ppm $C_3F_8$ | 113.3 | 6.9 |
| Trappist like | 5 pc | 10 days | 1 ppm $C_3F_8$ | 79.8 | 12.6 |
| Trappist like | 5 pc | 10 days | 10 ppm $C_3F_8$ | 129.4 | 11.9 |
| Trappist like | 5 pc | 10 days | 100 ppm $C_3F_8$ | 180.8 | 12.5 |
| Trappist like | 10 pc | 50 days | 1 ppm $C_3F_8$ | 16.3 | 2.6 |
| Trappist like | 10 pc | 50 days | 10 ppm $C_3F_8$ | 27.0 | 2.5 |
| Trappist like | 10 pc | 50 days | 100 ppm $C_3F_8$ | 40.5 | 2.4 |
| $NF_3$ | | | | | |
| Sun like | 5 pc | 10 days | 1 ppm $NF_3$ | 24.6 | 4.6 |
| Sun like | 5 pc | 10 days | 10 ppm $NF_3$ | 43.1 | 4.5 |
| Sun like | 5 pc | 10 days | 100 ppm $NF_3$ | 62.8 | 4.9 |
| Sun like | 10 pc | 50 days | 1 ppm $NF_3$ | 28.5 | 5.5 |
| Sun like | 10 pc | 50 days | 10 ppm $NF_3$ | 48.5 | 5.4 |
| Sun like | 10 pc | 50 days | 100 ppm $NF_3$ | 68.4 | 5.8 |
| K6V | 5 pc | 10 days | 1 ppm $NF_3$ | 57.2 | 11.2 |
| K6V | 5 pc | 10 days | 10 ppm $NF_3$ | 95.9 | 11.0 |
| K6V | 5 pc | 10 days | 100 ppm $NF_3$ | 133.1 | 12.0 |
| K6V | 10 pc | 50 days | 1 ppm $NF_3$ | 39.7 | 7.5 |
| K6V | 10 pc | 50 days | 10 ppm $NF_3$ | 68.1 | 7.2 |
| K6V | 10 pc | 50 days | 100 ppm $NF_3$ | 95.8 | 7.7 |
| Trappist like | 5 pc | 10 days | 1 ppm $NF_3$ | 49.2 | 8.3 |
| Trappist like | 5 pc | 10 days | 10 ppm $NF_3$ | 92.7 | 9.2 |
| Trappist like | 5 pc | 10 days | 100 ppm $NF_3$ | 142.0 | 11.6 |
| Trappist like | 10 pc | 50 days | 1 ppm $NF_3$ | 8.6 | 1.4 |
| Trappist like | 10 pc | 50 days | 10 ppm $NF_3$ | 17.4 | 1.6 |
| Trappist like | 10 pc | 50 days | 100 ppm $NF_3$ | 28.7 | 2.3 |
| $CF_4$ | | | | | |
| Sun like | 5 pc | 10 days | 1 ppm $CF_4$ | 4.2 | 1.3 |
| Sun like | 5 pc | 10 days | 10 ppm $CF_4$ | 11.3 | 2.4 |
| Sun like | 5 pc | 10 days | 100 ppm $CF_4$ | 25.2 | 3.0 |
| Sun like | 10 pc | 50 days | 1 ppm $CF_4$ | 3.9 | 1.3 |
| Sun like | 10 pc | 50 days | 10 ppm $CF_4$ | 10.4 | 2.1 |
| Sun like | 10 pc | 50 days | 100 ppm $CF_4$ | 24.1 | 2.8 |

Table 6
(Continued)

| Host Star | Distance | $T_{obs}$ | Feature | Band-int. ($\sigma$) | Max. Channel ($\sigma$) |
| --- | --- | --- | --- | --- | --- |
| K6V | 5 pc | 10 days | 1 ppm $CF_4$ | 6.4 | 1.9 |
| K6V | 5 pc | 10 days | 10 ppm $CF_4$ | 18.3 | 3.9 |
| K6V | 5 pc | 10 days | 100 ppm $CF_4$ | 42.8 | 5.2 |
| K6V | 10 pc | 50 days | 1 ppm $CF_4$ | 5.4 | 1.9 |
| K6V | 10 pc | 50 days | 10 ppm $CF_4$ | 13.5 | 2.3 |
| K6V | 10 pc | 50 days | 100 ppm $CF_4$ | 32.2 | 4.3 |
| Trappist like | 5 pc | 10 days | 1 ppm $CF_4$ | 15.4 | 7.0 |
| Trappist like | 5 pc | 10 days | 10 ppm $CF_4$ | 26.9 | 6.8 |
| Trappist like | 5 pc | 10 days | 100 ppm $CF_4$ | 63.6 | 9.0 |
| Trappist like | 10 pc | 50 days | 1 ppm $CF_4$ | 3.4 | 1.5 |
| Trappist like | 10 pc | 50 days | 10 ppm $CF_4$ | 6.1 | 1.4 |
| Trappist like | 10 pc | 50 days | 100 ppm $CF_4$ | 14.8 | 1.9 |

**Note.** The corresponding data for 1, 10, and 100 ppm combinations of $C_2F_6 + C_3F_8 + SF_6$ can be found in Table 5.

### ORCID iDs

Edward W. Schwieterman ⓘ https://orcid.org/0000-0002-2949-2163
Thomas J. Fauchez ⓘ https://orcid.org/0000-0002-5967-9631
Jacob Haqq-Misra ⓘ https://orcid.org/0000-0003-4346-2611
Ravi K. Kopparapu ⓘ https://orcid.org/0000-0002-5893-2471
Daniel Angerhausen ⓘ https://orcid.org/0000-0001-6138-8633
Daria Pidhorodetska ⓘ https://orcid.org/0000-0001-9771-7953
Michaela Leung ⓘ https://orcid.org/0000-0003-1906-5093
Evan L. Sneed ⓘ https://orcid.org/0000-0001-5290-1001
Elsa Ducrot ⓘ https://orcid.org/0000-0002-7008-6888